\documentclass[prl,aps,twocolumn]{revtex4}
\usepackage{amssymb}
\usepackage{graphicx}

\begin{document}
\bibliographystyle{revtex}
\title{Superconductivity in the charge-density-wave state
of the organic metal $\alpha$-(BEDT-TTF)$_2$KHg(SCN)$_4$}
\author{D.~Andres$^1$}
\author{M. V.~Kartsovnik$^1$}
\author{W.~Biberacher$^1$}
\author{K.~Neumaier$^1$}
\author{E. Schuberth$^1$}
\author{H. M\"uller$^2$}

\affiliation{$^1$Walther-Meissner-Institut, Bayerische Akademie der
Wissenschaften, D-85748 Garching, Germany}

\affiliation{$^2$European Synchrotron Radiation Facility, F-38043
Grenoble, France}

\begin{abstract}

The superconducting transition in the layered organic compound
$\alpha$-(BEDT-TTF)$_2$KHg(SCN)$_4$ has been studied in the two
hydrostatic pressure regimes where a charge-density wave is either
present or completely suppressed.
Within the charge-density-wave state the experimental
results reveal a network of
weakly coupled superconducting regions. This is especially seen in
a strong enhancement of the measured critical field and the
corresponding positive curvature of its temperature dependence.
Further, it is shown that on lowering the pressure into the
density-wave state traces of a superconducting phase already start
to appear at a much higher temperature.

\end{abstract}
\maketitle

\section{Introduction}

The organic metal \mbox{$\alpha$-(BEDT-TTF)$_2$KHg(SCN)$_4$} has
already raised great attention due to a variety of
novel physical phenomena found in its low-temperature
charge-density-wave (CDW) state \cite{wos96,kar96,bro95,singleton00}.
Of particular interest have been,
for example, new kinds of modulated CDW states existing in magnetic
fields above the paramagnetic limit \cite{bis98,chr00,and03,har04} and
phase
transitions induced by high magnetic fields due to a specific
interplay between the Pauli paramagnetic and orbital effects
\cite{and03,and01}. Apart from the high-field phenomena there
are other interesting properties, such as the coexistence/competition
of CDW and superconductivity which have not been thoroughly
addressed so far.

  Owing to a strongly anisotropic
electron system, the Fermi surface (FS) of this compound consists of
co-existing open sheets and cylinders \cite{mor91,rou96}. The slightly
warped
sheets correspond to a quasi-one-dimensional (Q1D) electron band. The
latter
emerges due to an enhanced electron transfer integral $t_a$ in the
crystallographic {\bf a}-direction between the organic
BEDT-TTF molecules resulting in a
chain-like coupling within the conducting {\bf a-c} plane
\cite{rou96}. At about 8~K there is a phase transition to the CDW
state \cite{bis98,chr00,ken97,fou03},
 in which these sheets of the FS become nested and the Q1D carriers are
 gapped. The system, however, keeps its metallic character
due to the second, quasi-two dimensional (Q2D) band.

Remarkably, the iso-structural salt
\mbox{$\alpha$-(BEDT-TTF)$_2$NH$_4$Hg(SCN)$_4$} (hereafter we refer
to both compounds as NH$_4$- and K-salt) does not undergo the density
wave transition but instead becomes superconducting (SC) at
$\approx$~1~K \cite{wang90,tan96}. The absence of a density wave is
interpreted to be due to a higher inter-
to intrachain-coupling ratio $t_c/t_a$ of the organic molecules within the
layers,
that strongly deteriorates the nesting conditions of the
open sheets of the FS \cite{kon03,mae01}. Moreover, it has been shown
\cite{mae01} that
by tuning the ratio of the lattice constants $c/a$ under uniaxial
strain a density wave can be even (i) induced in the NH$_4$-salt
and (ii) suppressed in the K-salt,
 a  SC state being stabilized at $\approx 1$~K.
Based on combined uniaxial strain measurements and band
structure calculations Kondo et al. \cite{kon03} have proposed
that the major contribution to superconductivity comes from
the Q1D band .

Similarly, hydrostatic pressure turns out to worsen the nesting
conditions in the K-salt \cite{and01}. The increase of the
interchain coupling leads to a decrease of the density wave
transition temperature,
 and at the pressure $P_c\approx 2.5$~kbar the density wave is completely
suppressed,  a normal metallic (NM) state being stabilized
\cite{and01}. Hydrostatic pressure studies \cite{and02} have
also revealed superconductivity in the K-salt but at temperatures
much lower than it was observed in the uniaxial strain experiments.
Remarkably, the superconductivity was
shown to persist over the whole pressure range studied, from 0 up to
4~kbar, i.e. it exists both in the NM and in the CDW
regimes. This offers a direct opportunity to study the influence
of a CDW on a SC system.

Basically, the SC pairing competes with the
density-wave instability for the FS \cite{ish98,gab02}. Therefore
one would expect the SC transition to be suppressed upon entering
the CDW region of the phase diagram since the Q1D carriers, which
are supposed to be responsible for superconductivity \cite{kon03},
become completely gapped below $P_c$. On the other hand,
it was predicted recently \cite{kuroki01,onari04} that
density-wave fluctuations can even stimulate the SC pairing
in the vicinity of the CDW ground state.

In this paper we present experimental studies of the SC transition
in the K-salt at different pressures, temperatures and magnetic
fields. We argue that below the critical pressure $P_c$ the SC
phase exists in the form of an array of weakly coupled small SC
regions or filaments embedded in the metallic CDW matrix.
Moreover, we show that the SC onset temperature becomes
drastically enhanced on lowering the pressure across the CDW/NM
boundary which is likely a sign of a nontrivial effect of the
CDW on the superconductivity in this compound.

\section{Experiment}

The main results presented in the paper were obtained from
interlayer resistance measurements using the standard four
probe geometry and a.c. measuring technique. Two samples,
hereafter referred to as samples $\#1$ and $\#2$, were measured
simultaneously. The samples had the
dimensions of $\sim 0.6\times 0.5\times 0.2$~mm$^3$ and
$1.0\times 0.3\times 0.05$~mm$^3$, respectively,
the smallest dimension being in the interlayer direction.
Additionally, measurements with the current applied along the
biggest dimension, i.e. nominally parallel to the layers, were
done on sample $\#2$. Of course, due to the extremely high
anisotropy of our compound this measured
``inplane'' resistance includes a mixture of the intra- and
interlayer components of the resistivity tensor \cite{kartsovnik04c}.
To minimize the influence of the interlayer component the
thinnest sample was chosen. The in- and interplane resistances
were measured in the same run by using the standard 6-probe
geometry (four
contacts were made to one of the biggest surfaces of the
plate like sample and two contacts to the opposite surface).
Thus, after comparing the measured in- and interplane resistances
we were able to  make reasonable conclusions about the
temperature dependence of the intralayer resistivity.

Hydrostatic pressure was applied using a conventional
berillium-copper clamp cell. The latter was mounted on a dilution
refrigerator allowing the sample to be cooled down to 20~mK. The
pressure value at low temperatures was determined from the
resistance of a calibrated manganin coil to an accuracy better than
$\pm$100~bar.

At the lowest temperatures, special care was taken to control and
minimize overheating due to the transport current and field-sweep
induced eddy currents. On measuring the interlayer resistance
with the applied current of 50~nA the overheating of the sample
was found to be $<$~5~mK at 20~mK. The sweep rates of the magnetic
field were chosen extremely low, $\approx 1$mT$/$min, so that
eddy currents had no visible effect on the sample temperature.

Further, since the SC properties are extremely sensitive to
magnetic fields, the superconducting magnet used in the experiment
was always carefully demagnetized before the measurements, so that
the remanent field was below 0.5~mT.

\section{Results and Discussion}

\subsection{Resistive SC transition at zero field}

\begin{figure}[t]
\center
\includegraphics[width=.95\linewidth]{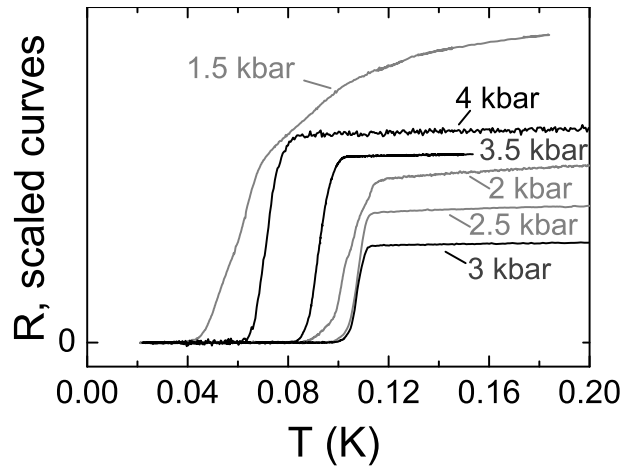}
\caption{Temperature sweeps of the interlayer resistance
of sample $\#~1$ at different pressures.
For clarity, each curve has a different resistance scale. At
$P>2.5$~kbar,
there are sharp transitions from the NM state to superconductivity.
Within the CDW state, $P \lesssim 2.5$~kbar, the superconducting
transitions are broadened and the zero-resistance temperature decreases.} \label{SCexample2}
\end{figure}

In Fig.~1 several temperature sweeps of the
interlayer resistance for sample $\# 1$ measured at different
pressures show the already reported behavior \cite{and02}. At $P=3$~kbar
the resistance exhibits a normal metallic behavior on cooling until at
110~mK a sharp SC transition ($\Delta T \approx 10$~mK) occurs. Above
$P_c\approx 2.5$~kbar the SC transition remains sharp and the critical
temperature $T_c$, defined as the midpoint of the transition, shows a
negative
pressure dependence of about -30~mK/kbar \cite{and02}. This value is
1-2 orders of magnitude lower than measured in other BEDT-TTF-based
superconductors, where a strong linear suppression of
superconductivity with hydrostatic pressure is
commonly observed \cite{ish98,lang03}.

Kondo et al. \cite{kon03} performed uni-axial strain experiments on
the NH$_4$-salt, with a
combined X-ray determination of the lattice parameters. Their tight
binding band structure calculation proposed the changes of the SC
transition temperature to be reasonably described
by the changing density of states (DOS) at the Fermi level within
the BCS model. However, they
mention that such a simple description fails as one approaches the
density wave state.
Under hydrostatic pressure,
the pressure dependence of $T_c$ in the K-compound is found to be an
order of
magnitude lower than observed \cite{cam95} in the NH$_4$-salt.
This is quite unusual: normally isostructural organic superconductors
with different anion layers display approximately the same
pressure dependence of $T_c$ \cite{ish98}. Thus, also
in the hydrostatic pressure case the proximity to the density-wave
instability in the K-salt seems to affect the SC transition in the
metallic
state. A direct comparison of the
SC properties between the two compounds may, therefore, be inappropriate.
Indeed, the value $dT_c/dP = -30$~mK$/$kbar is closer
to that observed in the Q1D TMTSF (or TMTTF)
based organic metals in which the SC state exists in the
hydrostatic pressure range right next to the spin-density-wave state
\cite{ish98}.
Obviously, in the vicinity of a density-wave transition a
detailed consideration of different carrier interactions, due to
which different instabilities of the metallic ground state compete
with each other, becomes necessary.

\begin{figure}[t]
\center
\includegraphics[width=.95\linewidth]{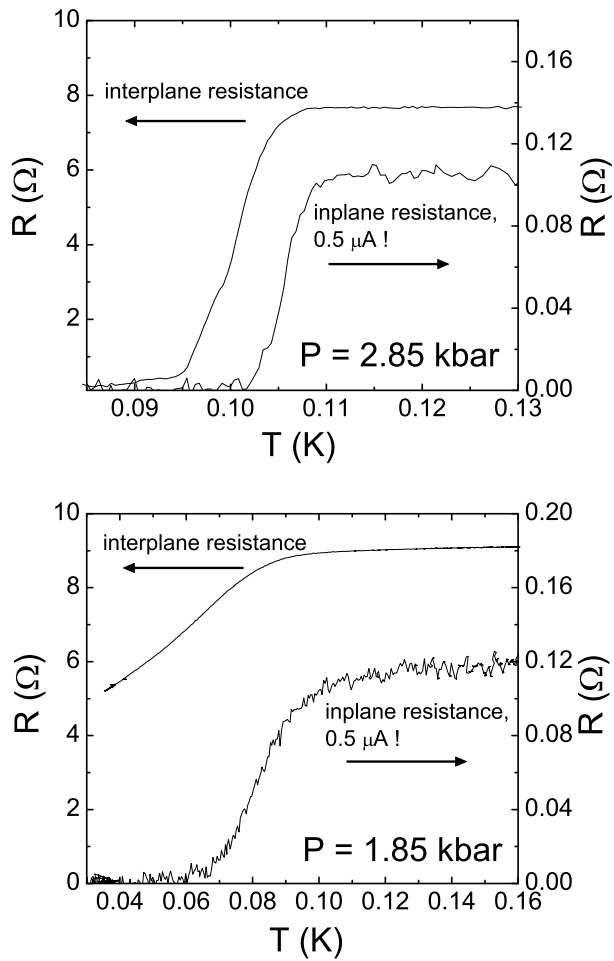}
\caption{Comparison of the in- and interplane resistances of
sample $\# 2$ at pressures above (upper graph) and below (lower graph)
the critical value $P_c$=2.5~kbar.} \label{SCinplane}
\end{figure}

As mentioned in the introduction, on entering the CDW state, i.e.
with lowering the pressure below $P_c$,
 the superconductivity does not vanish. At 2.5~kbar
$T_c$ remains at the value observed at
3~kbar, instead of further increasing, as would be
expected from an extrapolation from higher $P$.
With further decreasing the pressure, the transition
broadens and gets a kind of a step-like structure as
can be seen in Fig.~1. This leads
to a strong suppression of the temperature $T_0$ at which zero
resistance is reached; at ambient pressure the resistance does
not vanish down to 20~mK.
Thus, there is a clear effect of the CDW on the resistive SC
transition.
We note that the observed data are also very well in line
with the former proposal \cite{and01} of 2.5 kbar being about
the critical pressure $P_c$ for the complete suppression of the
CDW state.

\begin{figure}[t]
\center
\includegraphics[width=.75\linewidth]{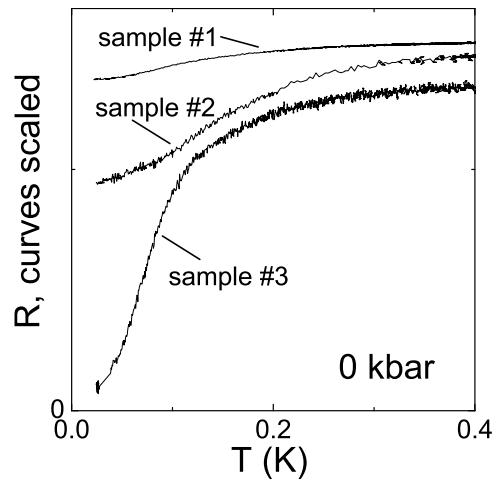}
\caption{Temperature dependence of the interlayer resistance at
         ambient pressure for three different samples.} \label{SCbsweeps}
\end{figure}

The overall behavior described above was also observed on sample $\#2$,
measured simultaneously. The superconducting
transition temperature, however, appears to be sample dependent. The
difference between the resistively measured transition temperatures
of samples $\#1$ and $\#2$ is approximately $10\%$ at $P~>$~2.5~kbar,
and becomes even stronger in the CDW state, at $P~\lesssim$~2.5~kbar.
This suggests the impact of the CDW on the superconductivity to be
also dependent on impurities or defects.
%
%


In Fig.~2 a comparison between the in- and interplane
resistances is shown for sample $\#2$ at pressures above and below
$P_c$.
Note that in order to measure the inplane resistance to a reasonable
accuracy the applied current had to be at least 0.5~$\mu$A. However,
despite this high current, that caused a small, $\sim$ 1-2~mK,
overheating at the transition temperature, it is seen
that the SC transition in the
plane occurs at a higher temperature in comparison to the
interlayer one. This difference in the transition temperatures
originates most likely from the layered character of superconductivity:
the SC ordering is first established within the layers whereas the
interlayer coherence develops at lower temperatures.
Such a scenario has
also been proposed for the NH$_4$-compound \cite{tan96}, where the
interlayer coherence length $\xi_\bot$ is found to be
smaller than the interlayer spacing of 20~\AA  \cite{ish98}.
This can also be assumed for the K-salt, since,
although $T_c$ is here an order of magnitude lower, the
in- to interplane anisotropy of the Fermi velocity is considerably higher
than the one in the NH$_4$ compound \cite{hanasaki01}.

At 1.85~kbar the inplane resistance is zero below 50-60~mK whereas
the interlayer transition does not vanish down to the lowest temperature.
A clear broadening of the inplane transition within the
CDW state is, however, also observed. We therefore presume that the
evolution, with pressure, of the SC transition in the intralayer
resistance is similar to that described above for
the interlayer resistance. This is supported by the previous
report by Ito et al. \cite{ito95} on the incomplete transition
in the inplane resistance at ambient pressure.


We now discuss a possible reason for broadening the SC
transition. First, we note that for all measured samples the transition
width is maximum at zero pressure and decreases as the pressure is
increased until the critical value $P_c$ is reached; at $P>P_c$ the
transition width is relatively small and approximately constant,
$\Delta T_c \approx 10$~mK. Thus, the broadening cannot be
ascribed to pressure inhomogeneity. Generally one
can think of phase fluctuations, typical of highly
anisotropic electron systems with small superfluid density, that
leads to a suppression of the bulk superconductivity \cite{eme95} as
has been observed in high $T_c$ superconductors \cite{tin96}.
However, in our system the SC transition temperature is of the order
of 100~mK. In this case, the zero-temperature
phase-stiffness of superconductivity
is high enough, so that effects of phase
fluctuations on $T_c$ are negligible \cite{eme95}.


\begin{figure}[t]
\includegraphics[width=.85\linewidth]{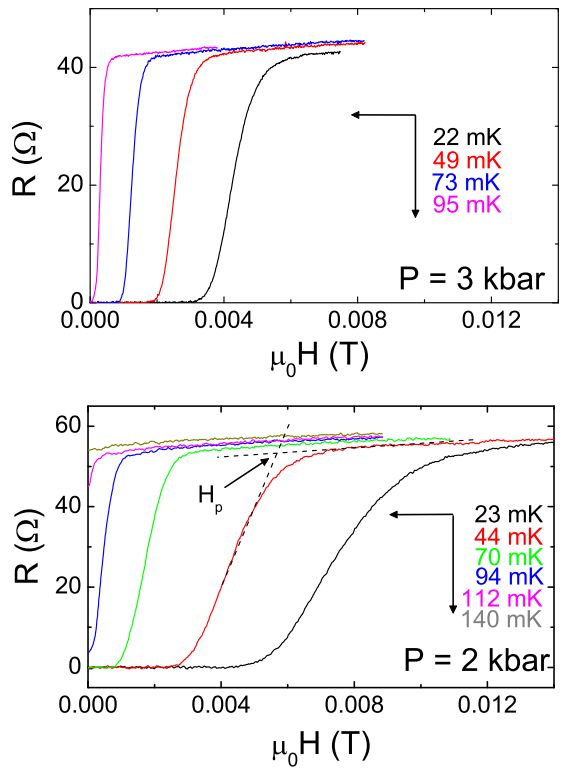}
\caption{Field dependence of the interlayer resistance
         of sample $\# 1$ at various
         constant temperatures for pressures above and below $P_c$.}
         \label{Bp}
\end{figure}

A clue to finding the real nature of the strongly broadened
resistive transition lies in a comparison of transport and
magnetization measurements. In Fig.~3 we show the
temperature dependence of the interlayer resistance for three
different samples at ambient pressure. As can be seen, sample $\#
3$ almost reaches zero resistance on cooling down to 20~mK,
reflecting the already mentioned sample dependence of the SC
transition \cite{and02}. This, however, does not mean that the
whole sample at lower temperatures is in the SC state. D.C.
magnetization measurements on the same sample made on a SQUID
magnetometer could not resolve any Meissner effect, even down to
6~mK.
Therefore, the zero resistance most likely originates from a network
of weakly coupled SC regions or filaments.
Thus, we suggest that the SC and CDW phases are separated in space.
This is also supported by theoretical predictions that a CDW leads
to a suppression of superconductivity \cite{gab02}. We consider an
inhomogeneous system of SC islands embedded in a metallic (actually
CDW) matrix to be more likely. The SC coherence, thus, develops
within the islands until at lower temperatures they couple to each
other via the proximity effect, providing a percolation network. At
ambient pressure the islands are strongly separated, so that a
completely coupled system does not exist at $T>20$~mK. A strong
broadening of the ``bulk'' SC transition is indeed known to exist in
a two dimensional array of SC islands which are embedded in a
metallic matrix \cite{res81,abr82}. After the islands become SC the
decrease of the resistance is determined by the growth of the normal
metallic coherence length on lowering the temperature, i.e. the
proximity effect.
Since we have no possibility at the moment to study the
magnetization under pressure, we cannot directly verify the absence
of the Meissner effect. However, as we shall see next, the
inhomogeneous nature of superconductivity under hydrostatic
pressure is supported by measurements of the SC transition in
magnetic fields.

\begin{figure}[t]
\includegraphics[width=.85\linewidth]{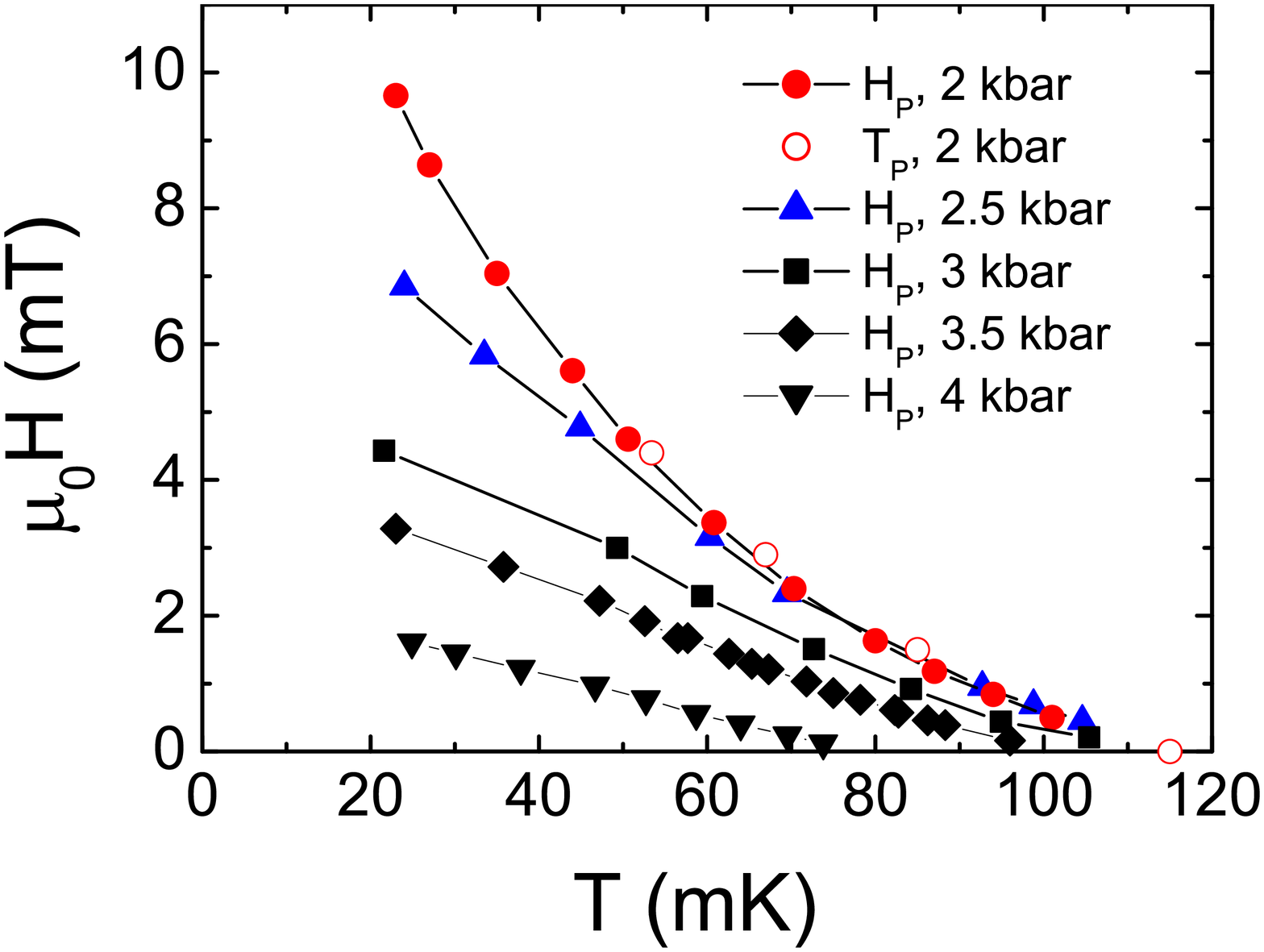}
\caption{Critical fields and temperatures determined at pressures
around the critical value $P_c$=2.5~kbar. Filled symbols are
obtained from field sweeps, see Fig.~4, and open circles from
temperature sweeps, see Fig.~7.}
\label{bpindep}
\end{figure}

\subsection{Magnetic field effect}

In Fig.~4 we show the magnetic field sweeps made on
sample~$\#1$, with the field directed perpendicular to the planes,
at different temperatures and two pressures, above and below the
critical value $P_c$. At zero field the transition
temperature at these two pressures is approximately the
same (see Fig.~1). While at 3~kbar
the transitions remain relatively sharp over the whole
temperature range, at 2~kbar they become somewhat broadened at
lower $T$. The critical fields $H_p$
determined as shown in the lower panel of Fig.~4 are
plotted in Fig.~5 for five
different pressures. At $P \geq 3$~kbar the critical field displays
a nearly linear dependence on temperature that can be expected
for coupled SC planes in the 3D limit \cite{tin96}. On entering
the CDW state, $H_p$ at low temperatures becomes dramatically
enhanced, leading to a pronounced positive curvature of its
temperature dependence as seen in the 2~kbar curve. It is important
to note that
this behavior does not depend on the way we
determine $H_p$. To illustrate this, Fig.~6 shows the
critical fields obtained by three different methods for two
different pressures, above and below $P_c$. Obviously, all
criteria lead to the same qualitative behavior.

In principle,
the positive curvature of $H_p$ might be related to the
melting of the superconducting vortex lattice. However, this can
be ruled out by looking at the temperature dependence of the
interlayer resistance at different constant magnetic fields that
is shown in Fig.~7.
The fact that at 2 kbar the resistive transition in the
temperature sweep does not broaden with applying a magnetic
field rules out any considerable flux flow effect.
It would, anyway, be surprising if the vortex motion were important
in a material with such a low $T_c$.


\begin{figure}[t]
\center
\includegraphics[width=.80\linewidth]{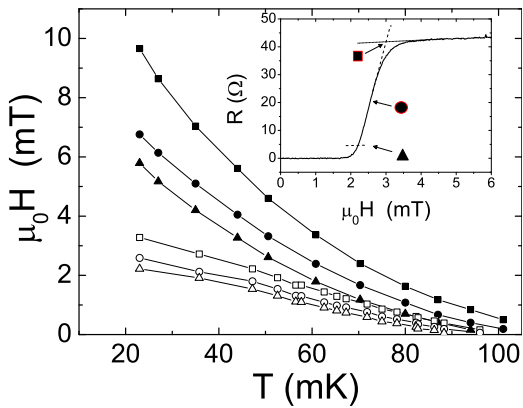}
\caption{Critical fields at $P$=2~kbar (solid symbols)
         and $P$=3.5~kbar (open symbols) determined from the resistive
         transition in the field sweeps using different criteria. The inset
         shows how the criteria are chosen: the onset (squares), the
         inflection point (circles), and the end of the SC transition
         where the resistance is $\approx 10 \%$ of the normal metallic
         value (triangles).} \label{Tp}
\end{figure}

The upper critical field $H_{c2}$ is generally known to be enhanced
in a superconductor if at least one of the dimensions perpendicular
to the field direction becomes less than the coherence length
\cite{tin96}. A dimensional crossover with lowering $T$
 then also leads to a strong positive curvature of
the upper critical field. A similar scenario might also occur
in our compound. This means that there's a possibility that
the size of the superconducting
regions within the plane becomes less
than the coherence length.
However, the field $H_p$ determined from the resistive transition
does not necessarily match the upper critical field
$H_{c2}$ at $P<P_c$. As was argued above, the superconductivity
is most likely inhomogeneous in this pressure range. Therefore,
the resistive transition may be largely determined by the
coupling between the randomly separated SC islands rather
than by $H_{c2}$ inside the islands.
This means that not only the value of $H_p$ defined above
can differ from the real $H_{c2}$ but also its temperature
dependence.
Although an exact theoretical description of the
resistive transition of a proximity coupled random array of SC
islands
in a magnetic field still has to be worked out, a comparison to
existing inhomogeneous superconductors shows that a strong positive
curvature of $H_p$ can be expected.

As an example one can mention polymeric sulfur nitride (SN)$_x$, a
compound that consists of bundles of SC filaments. For a magnetic
field applied perpendicular to the fiber axis the temperature
dependence of the resistive transition was shown to exhibit a
positive curvature \cite{aze76}.
Another, and probably more relevant example is the well known CDW
compound NbSe$_3$. It has been reported \cite{bri81} that within the
CDW state of NbSe$_3$ a small fraction of the sample becomes SC  and
it has been proposed to emerge within the boundaries of CDW domain
walls, where the CDW order parameter is supposed to become zero.
This would then indeed be a system of SC regions separated by the
metallic CDW phase similarly to our present case. At higher
pressures the CDW gap becomes smaller and the domain wall fraction,
where ungapped Q1D electrons exist, is expected to become bigger.
Moreover, a strong sample dependence of the SC properties would not
be surprising in such a model, since crystal defects or impurities
very likely affect the domain structure. Whether such a domain
structure really exists in the title compound we cannot judge from
our data, but the similarities between both compounds with respect
to their SC properties suggest the nature of the critical field
behavior to be the same. The possibility of domains within a
Q1D CDW system has indeed been
predicted \cite{gor83}. Furthermore,
Gor'kov et al. mention that the superconductivity would
be expected  to survive in
the domain walls perpendicular to the conducting chain direction
\cite{gor83,gor05}.

\begin{figure}[t]
 \center
\includegraphics[width=.75\linewidth]{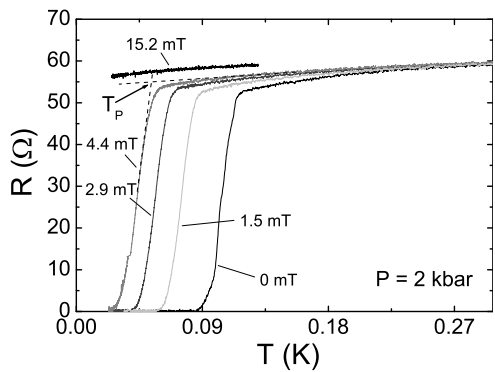}
\caption{Temperature dependent interlayer resistance at different
         constant magnetic fields, at $P=2$~kbar.}
\label{hightc}
\end{figure}


Noteworthy, there might exist a narrow
pressure region in the vicinity of $P_c$, in which the system becomes
inhomogeneous, irrespective of the CDW domain structure \cite{vuletic02}.
Such an
inhomogeneous system, associated with a first order phase
transition, was also shown to have an enhanced SC upper critical
field\cite{lee02} in the spin density wave compound (TMTSF)$_2$PF$_6$.

\subsection{Enhanced SC onset temperature}

\begin{figure}[t]
\center
\includegraphics[width=.85\linewidth]{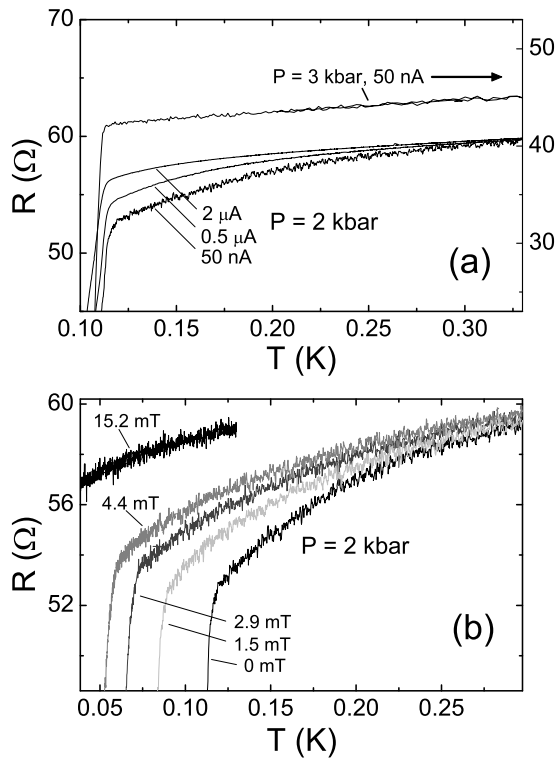}
\caption{Fig.~8. Within the CDW state ($P=2$~kbar) the decrease of the
         interlayer resistance accelerates at much higher temperatures than
         in the NM state ($P=3$~kbar). This decrease strongly depends on the
         level of the applied current (a) and magnetic field perpendicular to
         the layers (b).} \label{PTdiag2}


\end{figure}

Besides the broadening of the main SC transition, all temperature sweeps
at pressures $\leq 2.5$~kbar show an unusually strong decrease
(negative curvature) of the resistance in a remarkably wide
temperature range well above the $T_c$ value that would be expected from
its linear extrapolation from $P>P_c$. Fig.~8
shows, in an enlarged scale, the resistance of sample $\#1$
at 2~kbar, at temperatures right above the main transition,
which is still rather sharp at this pressure. For comparison, the
3~kbar resistance
is also shown in the upper panel. As can be seen from the figure,
the decrease of the resistance strongly depends on the level of the
applied current and field. With increasing the current or field
the resistance decrease becomes suppressed.
Note that the main transition shifts only slightly at higher
currents in Fig.~8.
Therefore, effects of overheating  can be neglected.

The present data manifest that traces of superconductivity,
occupying a small fraction of the crystal volume, exist
already at much higher temperatures. The described behavior was
found throughout the entire CDW pressure range. The onset
temperature of superconductivity is $\approx 0.22$~K at 2.5~kbar
and 0.30~K at 2~kbar and 0~kbar.
These findings were reproduced on several samples. They are also
consistent with the ambient pressure results of
Ito et al. \cite{ito95}

\begin{figure}[t]
\center
\includegraphics[width=.85\linewidth]{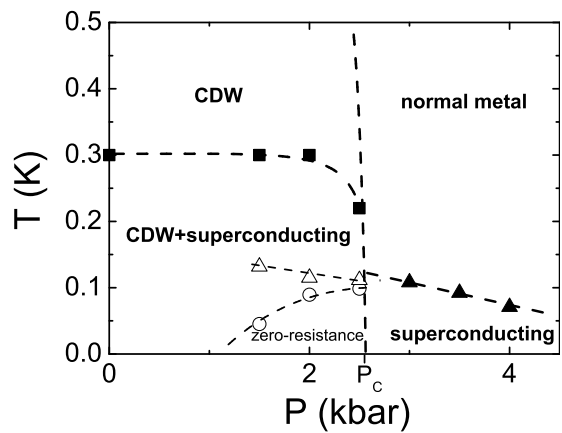}
\caption{Proposed {\it P-T} phase diagram. Filled symbols show the
phase transitions between different states. Open symbols mark the
onset and zero-resistance temperatures of the broadened main SC
transition in the CDW state. The lines are guides for the eye.
} \label{figure9}

\end{figure}

 By contrast to the CDW pressure region, in the NM state such an
  accelerated decrease of the resistance above the bulk SC transition
  has not been detected (see the 3~kbar curve in Fig.~8).
  Hence, we conclude that the dramatic increase of the SC onset
  temperature is a consequence of entering the CDW region of the
  phase diagram.
The whole $P$--$T$ phase diagram including all phases must,
therefore, look as depicted in Fig.~9.
Since the SC transition is sharp above $P_c$ but becomes
broadened in the CDW region, we take here the midpoint for the NM$/$SC
transition (filled triangles in Fig.~9) and the onset and
zero-resistance temperatures for the main SC transition in the CDW state
(open triangles and circles, respectively). The onset temperature of
small
SC regions in the CDW state (filled squares) is determined by the
inflection point in the temperature dependent resistance.
Obviously, there is an extended range in the $P$--$T$ phase diagram
that includes both ground states, superconductivity and density wave.

 If the superconductivity is indeed spatially restricted to the CDW
domain boundaries, as suggested above, one can understand, why the
CDW does not completely suppress the SC state, in contradiction to
what has been theoretically proposed \cite{gab02}. This will,
however, not explain the enhanced SC onset temperature. In principle,
in the model above one would still expect the opposite effect,
namely that the SC island has a reduced onset temperature due to
the proximity effect.
On the other hand, we do not know in what way the superconductivity,
located in the domain boundaries where the order parameter
of the density wave reaches zero, is influenced by the CDW neighborhood.
 An interesting scenario to consider would be an additional
 stimulation of superconductivity in the CDW domain walls, such as,
 for example, a charge-fluctuation mediated pairing
 \cite{kuroki01,merino01,onari04}. More investigations on this topic are
 highly desirable.

\section{Conclusion}

 In conclusion, pronounced differences in the
superconducting properties are observed between the CDW and the NM
pressure regions. The determined phase diagram further confirms that
$P_c\approx 2.5$~kbar is the critical pressure at which the CDW state
becomes completely suppressed.
Below $P_c$, the broadening of the resistive SC transitions, the
absence of the Meissner effect as well as the pronounced enhancement
and positive curvature of the critical magnetic field point to the
formation of a network of coupled SC regions embedded in the CDW
matrix. We propose that the superconductivity is located within CDW
domain walls. Furthermore, it is found that traces of a SC phase
exist in the CDW region already at temperatures much higher than
expected from the NM state. The origin of this remarkable and
unexpected expansion of the SC temperature range remains at present
one of the most intriguing questions.

\section*{Acknowledgements}

The work is partially supported by the DFG-RFBR grant 436 RUS 113/592.

\end{document}